\documentclass[pra,showpacs,twocolumn]{revtex4}
\usepackage{bm,graphicx,amsmath}

\begin{document}

\title{Rabi oscillations in a quantum dot-cavity system coupled to a non-zero temperature phonon bath}
\author{Jonas Larson$^1$ and H\'ector Moya-Cessa$^{2}$}
\address{$^{1}$ICFO-Institut de Ci\`{e}ncies Fot\`{o}niques, E-08860 Castelldefels, Barcelona, Spain \\ $^{2}$INAOE,
Coordinaci\'on de Optica, Apdo. Postal 51 y 216, 72000 Puebla,
Pue., Mexico}
\date{\today}

\begin{abstract}
We study a quantum dot strongly coupled to a single high-finesse
optical microcavity mode. We use a rotating wave approximation
method, commonly used in ion-laser interactions, tegether with the
Lamb-Dicke approximation to obtain an analytic solution of this
problem. The decay of Rabi oscillations because of the
electron-phonon coupling are studied at arbitrary temperature and
analytical expressions for the collapse and revival times are
presented. Analyses without the rotating wave approximation are
presented by means of investigating the energy spectrum.
\end{abstract}

 \maketitle

\section{Introduction}
Semiconductor quantum dots (QD) have emerged as promising candidates
for studying quantum optical phenomena \cite{feng}. In particular,
cavity quantum electrodynamics (CQED) effects can be investigated
using a single QD embedded inside a photonic nano-structure
\cite{kiraz}. One of the most fundamental systems in CQED is an atom
interacting with a quantized field \cite{plk}, such a system has
been an invaluable tool to understand quantum phenomena
\cite{cqedphen} as well as for considerations on its applications to
realize quantum information \cite{cqedinf}. Similar systems (in the
sense of their treatment, applications, etc.) like trapped ions
interacting with lasers \cite{trap} have shown to be an alternative
to develop techniques for quantum information processing
\cite{ioninf} and the study of fundamental effects \cite{ionfun}.
Recent developments in semiconductor nano-technology have shown that
excitons in QD constitute yet another two-level system for CQED
considerations, and several successful experiments have been carried
out \cite{wallraff}, see also the referee \cite{cqd}. Contrary to the atom-field interaction where
dissipative effects can be fairly overlooked provided the coupling
strength between atom and field is sufficiently large compared to
the dissipation rate due to cavity losses, the physics of a quantum
dot microcavity is enriched by the presence of electro-electron and
electron-phonon interactions. Thus, decoherence due to phonons may
imply fundamental limitations to quantum information processing on
quantum dot CQED \cite{wuerger}. Here, we would like to analyze the
effects of electron-phonon interactions on electron-hole-photon Rabi
oscillations in cavity QED.

As in Ref. \cite{wilson,chinese}, we will not apply the Born-Markov
approximation \cite{bm}, but will use a different technique for
solving this problem. In particular we will make use of techniques
commonly applied in ion-laser interactions. It relies on the
rotating  wave approximation RWA, and within this and the Lamb-Dicke
approximation the Hamiltonian becomes diagonal with respect to the
phonon subsystem, which is shown in sec. \ref{sec2}. In the validity
regime of the RWA, the decoherence effect on the inversion, due to
the phonon bath, is analyzed in sec. \ref{sec3} and  analytical
expressions for the collapse and revival times are given. The zero
temperature situation has been investigated in \cite{chinese}, while
here we study the effects due to zero as well as non-zero
temperatures (causing the collapse of the revivals), and also how
different phonon mode structures affect the decoherence. A different
method to treat the non-zero temperature case was discussed in
\cite{non-zero}, where the collapse time is obtained numerically.
The dynamics beyond the rotating wave approximation, shortly studied
here in sec. \ref{sec4}, becomes highly complex as can be seen from
the energy spectrum. In what sense the phonon decoherence could be
used as a possible resource for various applications is briefly
mentioned in the concluding remarks.

\section{The model}\label{sec2}
We assume a simple two-level model for the electronic degrees of
freedom of the QD, consisting of its ground state $|g\rangle$ and
the lowest energy electron-hole (exciton) state $|e\rangle$, with
the Hamiltonian  \cite{wilson} ($\hbar=1$)
\begin{equation}
\begin{array}{lll}
\hat{H} & = & \displaystyle{\omega_{eg} \hat{\sigma}_{ee}+
\omega\hat{a}^{\dagger}\hat{a}+g(\hat{a}^{\dagger}\hat{\sigma}_-+\hat{\sigma}_+\hat{a})} \\ \\
& & \displaystyle{+ \hat{\sigma}_{ee} \sum_k\lambda_k
(\hat{b}_k^{\dagger}+\hat{b}_k)+\sum_k\omega_k
\hat{b}_k^{\dagger}\hat{b}_k}.
\end{array}
\end{equation}
Here $\hat{\sigma}_+ =|e\rangle \langle g|$, $\hat{a}$ and
$\hat{b}_k$ are the annihilation operators for the cavity mode and
the $k$th phonon mode, respectively. By transforming to a rotating
frame, with frequency $\omega$
\begin{equation}
\hat{\cal{H}}\! =\! \Delta \hat{\sigma}_{ee}
+g(\hat{a}^{\dagger}\hat{\sigma}_-+\hat{\sigma}_+\hat{a}) +
\hat{\sigma}_{ee}\!\! \sum_k\!\lambda_k (\hat{b}_k^{\dagger}+\hat{b}_k)
+\!\sum_k\!\omega_k \hat{b}_k^{\dagger}\hat{b}_k
\end{equation}
where $\Delta=\omega_{eg}-\omega$ is the detuning. The
transformation \cite{wuerger,duke}
\begin{equation}
\hat{T} = \prod_k
e^{\hat{\sigma}_{ee}\frac{\lambda_k}{\omega_k}(\hat{b}_k^{\dagger}-\hat{b}_k)}
\end{equation}
is used to obtain the Hamiltonian
$\hat{\cal{H}}_T=\hat{T}\hat{\cal{H}}\hat{T}^{\dagger}$
\begin{eqnarray}
\hat{\cal{H}}_T &=& \left(\Delta-\Delta_{\eta}\right)
\hat{\sigma}_{ee}+\sum_k\omega_k \hat{b}_k^{\dagger}\hat{b}_k
\\ &+&
g\left[\hat{a}^{\dagger}\hat{\sigma}_-\prod_k\hat{D}_{\hat{b}}\left(\eta_k\right)
+\hat{\sigma}_+\hat{a}\prod_k\hat{D}_{\hat{b}}^{\dagger}\left(\eta_k\right)\right]
 \nonumber ,
\end{eqnarray}
with $\eta_k=\lambda_k/\omega_k$, $\Delta_{\eta}=\sum_k\omega_k
\eta_k^2$ is the so-called polaron shift \cite{duke} and
$\hat{D}_{\hat{b}}(\eta_k)=\exp\left(\eta_k\hat{b}^\dagger-\eta_k^*\hat{b}\right)$.
For simplicity we look at the case $\Delta=\Delta_{\eta}$ to obtain
\begin{equation}\label{noRWA}
\hat{\cal{H}}_T\!
=\!g\!\left[\hat{a}^{\dagger}\hat{\sigma}_-\!\prod_k\!\hat{D}_{\hat{b}}\left(\eta_k\right)
+\hat{\sigma}_+\hat{a}\!\prod_k\!\hat{D}_{\hat{b}}^{\dagger}\left(\eta_k\right)\right]
 +\!\sum_k\omega_k
\hat{b}_k^{\dagger}\hat{b}_k.
\end{equation}
Now by transforming to an interaction picture one obtains
\begin{equation}\label{HO}
\hat{H}_I\!
=\!g\!\left[\hat{a}^{\dagger}\hat{\sigma}_-\!\prod_k\!\hat{D}_{\hat{b}}\!\left(\eta_k
e^{-i\omega_k t}\right)
+\hat{\sigma}_+\hat{a}\!\prod_k\!\hat{D}_{\hat{b}}^{\dagger}\!\left(\eta_k
e^{-i\omega_k t}\right)\!\right].
\end{equation}
The idea is now to apply the RWA on the above Hamiltonian,
equivalent to neglecting all rapidly oscillating terms. However, a
closer look at (\ref{HO}) indicates that such a procedure is not
straightforward. Normally $\omega_k=\omega_0k$, $k=1,2,3,...$ and
expanding the above exponents give several time-independent cross
terms leading to a complicated expression. However, it is know that
the RWA is highly related to the Lamb-Dicke approximation
\cite{dong}. In the Lamb-Dicke approximation it is assumed that the
variations of the exponents are smaller compared to the
characteristic length of the phonon harmonic oscillators, typically
$\eta_k\ll1$. One finds that in order to fulfill the RWA one
normally needs to be in the Lamb-Dicke regime \cite{dong}. This
observation helps us considerably in approximating the above
Hamiltonian, since now the cross time-dependent terms arising from
different $k$'s vanish as we assume $\eta_k\ll1$. In this case we
can perform the RWA separately on each individual exponent to find
\begin{equation}\label{rwaH}
\hat{H}_{rwa} =\tilde{g}\left[\hat{a}^{\dagger}\hat{\sigma}_-\prod_k
L_{\hat{n}_k}(\eta_k^2) +\hat{\sigma}_+\hat{a}\prod_k
L_{\hat{n}_k}(\eta_k^2)\right]
\end{equation}
where $L_{\hat{n}_k}$ are the Laguerre polynomials of order
$\hat{n}_k$ \cite{wine,moya} and $\tilde{g}=g\exp(\xi/2)\equiv
g\exp\left(-\sum_k\eta_k^2/2\right)$ is a rescaled Rabi vacuum
frequency. We emphasize that to be consistent with the RWA and
Lamb-Dicke approximation, only terms up to $\eta_k^{2}$ should be
considered when the Laguerre polynomials are expanded in the small
parameter $\eta^2$. This will be done in the following section. The
parameter $\xi$ is sometimes referred to as the Huang-Rhys factor
\cite{hr}, and it is usually very small, $\xi\ll1$,
\cite{hrsmall,rabidot} but it can become much larger, $\xi\sim1$,
\cite{hrbig} and then our analysis would break down. The above
equation is readily solvable, finding the evolution operator as
\begin{equation}
\hat{U}= \hat{U}_{ee}\hat{\sigma}_{ee} +
\hat{U}_{gg}\hat{\sigma}_{gg} + \hat{U}_{eg}\sigma_- +
\hat{U}_{ge}\sigma_+ ,
\end{equation}
where
\begin{equation}
\hat{U}_{ee}(t;\hat{n}_k) = \cos\hat{\Omega}_{k,\hat{n}+1}t,
\end{equation}
\begin{equation}
\hat{U}_{ge}(t;\hat{n}_k) = -i\epsilon\hat{a}
\frac{\sin\hat{\Omega}_{k,\hat{n}}t}{\hat{\Omega }_{k,\hat{n}}},
\end{equation}
\begin{equation}
\hat{U}_{eg}(t;\hat{n}_k) = -i\epsilon\hat{a}^{\dagger}
\frac{\sin\hat{\Omega}_{k,\hat{n}+1}t} {\hat{\Omega}_{k,\hat{n}+1}},
\end{equation}
and
\begin{equation}
\hat{U}_{gg}(t;\hat{n}_k) = \cos\hat{\Omega}_{k,\hat{n}}t,
\end{equation}
with
\begin{equation}
\hat{\Omega}_{\hat{n}_k} =\hat{\epsilon}\sqrt{\hat{n}_k
}=\tilde{g}L_{\hat{n}_k}(\eta_k^2)\sqrt{\hat{n}_k}.
\end{equation}

\section{Dynamics}\label{sec3}
Having the evolution operator, we can in principle calculate any
properties we want, in particular we look at the Rabi oscillations
for the two-levels system This quantity has as well been studied
experimentally in quantum dot systems \cite{rabidot}. By means of
the inversion operator $\sigma_z=|e\rangle\langle
e|-|g\rangle\langle g|$
\begin{equation}
W(t)=Tr\{\rho(t)\sigma_z\}
\end{equation}
where $\rho(t)=\hat{U}\rho(0)\hat{U}^\dagger$ and $\rho(0)$ is the
initial density matrix,  Rabi oscillations will be analyzed. This
initial state is chosen as the fully separable state
\begin{equation}
\rho(0)=\prod_k\rho_k(0)|0,e\rangle\langle0,e|
\end{equation}
with the QD excited, the cavity mode in vacuum and the phonon modes are all
assumed to be in a thermal distribution
\begin{equation}
\rho_k(0)\!=\! \sum_{n_k=0}^\infty\!|c_{n_k}|^2|n_k\rangle\langle n_k|\!=\!\sum_{n_k=0}^{\infty}\!\frac{\bar{n}_k^{n_k}}{(\bar{n}_k+1)^{n_k+1}}|n_k\rangle\langle
n_k|.
\end{equation}
The inversion can be written explicitly as
\begin{equation}\label{atinv}
W(t)=\sum_{\{n_k\}}\prod_k|c_{n_k}|^2\cos\left(2\tilde{g}\prod_{k'}L_{n_{k'}}(\eta_{k'}^2)t\right),
\end{equation}
where the summation goes from 0 to $\infty$ and runs over all modes.
For $t=0$ we have $W(0)=1$ as expected, while for $t\neq0$ the sum
will in general differ from 1. Interestingly we note that at zero
temperature, $T=0$, $|c_{n_k}|=\delta_{n_k,0}$ for all $k$ and thus;
at absolute zero temperature the system Rabi oscillations are intact
but with a rescaled frequency \cite{chinese}. The expression for the
inversion can be simplified by taking into account that $\eta\ll1$
and we therefore expand the Laguerre polynomials
\begin{equation}\label{Lexp}
L_n(\eta^2)=\!\sum_{m=0}^n\left(\begin{array}{c}
n \\ m
\end{array}\right)\frac{\eta^{2m}}{m!}=1\!-\!n\eta^2\!+\!\frac{n(n-1)}{4}\eta^4\!+\!...\,.
\end{equation}
Keeping only zeroth and first order terms in $\eta_k^2$ in the
assumed Lamb-Dicke regime, the inversion can be written, after some
algebra, as
\begin{equation}\label{appinv}
\begin{array}{lll}
W(t) & = & \displaystyle{\cos(2\tilde{g}t){\mathcal Re}\left(\prod_k\frac{1}{\bar{n}_k+1-\bar{n}_k\mathrm{e}^{-i2\tilde{g}t\eta_k^2}}\right)} \\ \\
& & \displaystyle{-\sin(2\tilde{g}t){\mathcal Im}\left(\prod_k\frac{1}{\bar{n}_k+1-\bar{n}_k\mathrm{e}^{-i2\tilde{g}t\eta_k^2}}\right)}.
\end{array}
\end{equation}
In the following we use dimensionless variables such that the
quantum dot-cavity coupling $g=1$, but  keep it in the formulas
for clarity. There are some special cases worth studying
separately.

\subsection{Single phonon mode}
\begin{figure}[ht]
\begin{center}
\includegraphics[width=8cm]{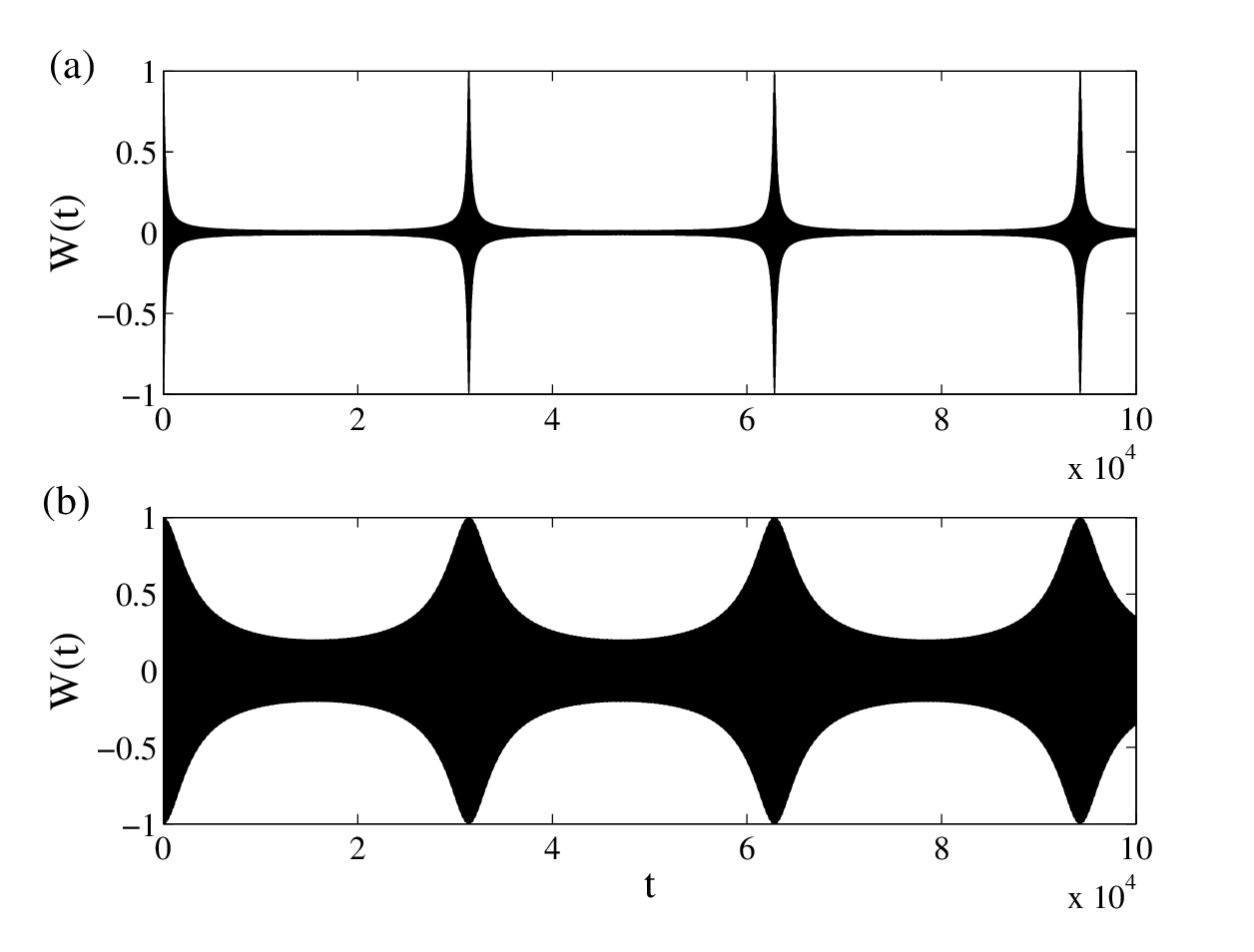}
\caption{\label{fig1}  The atomic inversion $W(t)$ as a function
of time $t$ for the single mode bath.  The dimensionless
parameters are $\eta=0.01$ and $\bar{n}=40$ (a)  and $\eta=0.01$
and $\bar{n}=2$ (b). }
\end{center}
\end{figure}

In the simplest case consisting of a single phonon mode
characterized by $\eta$ and $\bar{n}$, the inversion
(\ref{appinv}) simplifies to
\begin{equation}\label{appinv1m}
W(t)=\frac{(\bar{n}+1)\cos(2\tilde{g}t)-\bar{n}\cos\left[2\tilde{g}t(1-\eta^2)\right]}{(\bar{n}+1)^2+\bar{n}^2-2\bar{n}(\bar{n}+1)\cos(2\tilde{g}t\eta^2)}.
\end{equation}
The second term oscillates with a slightly shifted frequency
compared to the first term  causing a collapse of the inversion.
When the two competing terms return back in phase, at times
\begin{equation}
t_{rev}^{(1)}=k\frac{\pi}{\tilde{g}\eta^2},\hspace{1cm}k=1,2,3,...\,,
\end{equation}
inversion revivals occur. These are perfect within the small
$\eta^2$ expansion. For short times $t$, the  inversion may be
further approximated to give
\begin{equation}
W(t)=\frac{\cos(2\tilde{g}t)}{1+8\bar{n}(\bar{n}+1)\tilde{g}^2\eta^4t^2},\hspace{1cm}2\tilde{g}t\eta^2\ll1,
\end{equation}
and we conclude that the envelope function, determining the
collapse time, is a Lorentzian with width
\begin{equation}
t_{col}^{(1)}=\frac{1}{\sqrt{8\bar{n}(\bar{n}+1)}\tilde{g}\eta^2}.
\end{equation}
In fig. \ref{fig1} we display two different examples of the atomic
inversion (\ref{atinv}). The upper plot (a) has a large average
number of phonons; $\bar{n}=40$ and $\eta=0.01$. In the lower plot
(B) the number of phonons is instead $\bar{n}=2$ and again
$\eta=0.01$. We can conclude that our results confirm that a lower
temperature of the reservoir clearly increases the collapse time.

\subsection{$N$ identical phonon modes}
By studying several identical phonon modes one may see the effect
of multi modes in a simple analytic way. Instead of approaching eq. (\ref{atinv}) we go back to eq. (\ref{noRWA}) and let $\eta=\eta_k$ and $\omega=\omega_k$ for all $k$. Let us introduce a $N\times N$ dimensional unitary operator connecting the boson operators $\hat{b}_1,\hat{b}_2,...,\hat{b}_N$ with new ones $\hat{c}_1,\hat{c}_2,...,\hat{c}_N$ such that
\begin{equation}
\hat{c}_1=\frac{1}{\sqrt{N}}\sum_{k=1}^N\hat{b}_k.
\end{equation}
The unitary transformed Hamiltonian becomes
\begin{equation}\label{noRWA2}
\hat{\cal{H}'}_T\!
=\!g\!\left[\hat{a}^{\dagger}\hat{\sigma}_-\!\hat{D}_{\hat{c}_1}\left(\eta\sqrt{N}\right)
+\hat{\sigma}_+\hat{a}\!\hat{D}_{\hat{c}_1}^{\dagger}\left(\eta\sqrt{N}\right)\right]
 +\!\omega\!\!\sum_i
\hat{c}_i^{\dagger}\hat{c}_i.
\end{equation}
Thus, the problem relaxes to the single mode case with the scaled
Lamb-Dicke parameter $\eta\rightarrow\eta\sqrt{N}$.

\subsection{$N$ different phonon modes}
\begin{figure}[h]
\begin{center}
\includegraphics[width=8cm]{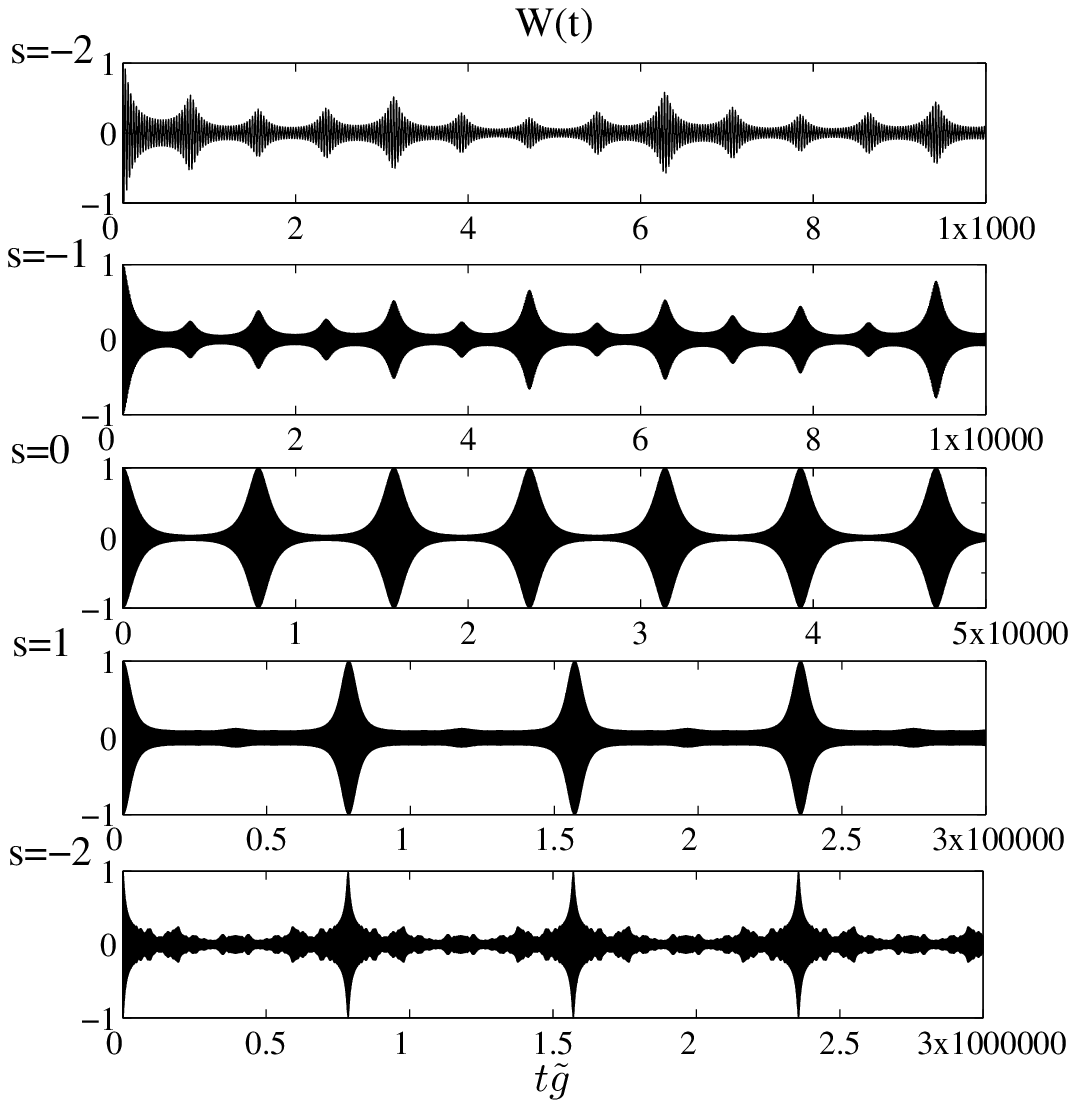}
\caption{\label{fig3} Atomic inversion for
different phonon modes obtained from the small $\eta^2$ expansion
(\ref{appinv}). The plots show respectively from top to bottom the
cases with powers $s=-2,-1,0,1,2$ of eq. (\ref{powercoup}). The
other parameters are the same for all five cases; $T=0.2$,
$\kappa=0.0004$ and $\omega_0=0.1$. }
\end{center}
\end{figure}

In a more realistic model the phonon bath consists of
non-identical modes, and  depending on the model studied one has
different spectral functions
$J(\omega)=\sum_k\lambda_k^2\delta(\omega-\omega_k)$ \cite{mahan}.
For the frequencies of interest here, $J(\omega)$ has a simple
power law behaviour \cite{power}, resulting in a
\begin{equation}\label{powercoup}
\eta_k^2=\kappa\omega_k^s,\hspace{1cm}s=...,-2,-1,0,1,2,...\,,
\end{equation}
where $\kappa$ is a constant. The power $s$ depends on matter
properties, for example;  {\it ohmic damping}
$J(\omega)\propto\omega$, {\it phonon damping}
$J(\omega)\propto\omega^3$ or {\it impurity damping}
$J(\omega)\propto\omega^5$, but it also depends on system
dimensions. We take $\omega_k=\omega_0k$, $k=1,2,3,...$ such that
$\omega_0$ determines the frequency spacing, while the thermal
phonon distributions are determined from the average phonon
numbers
\begin{equation}
 \bar{n}_k=\left[\exp\left(\frac{\omega_k}{\tilde{T}}\right)-1\right]^{-1},
\end{equation}
for some scaled temperature $\tilde{T}$. Often a frequency
``cut-off'' is introduced for the spectral function, but because the
vacuum modes do not affect the system dynamics in our case, such a
cut-off is not needed. Defining
$\gamma_k=8\bar{n}_k(\bar{n}_k+1)\tilde{g}^2\eta_k^4$, and using the
same arguments as above the collapse time in the small $\eta^2$
expansion is given by
\begin{equation}
 \prod_k\left(1+\gamma_kt_{col}^2\right)=2.
\end{equation}
By further introducing
\begin{equation}\label{parasum}
\begin{array}{l}
 r_k=\sqrt{(\bar{n}_k+1)^2+\bar{n}_k^2-2\bar{n}_k(\bar{n}_k+1)\cos(2\tilde{g}t\eta_k^2)},
\\ \\
\displaystyle{\theta_k=\arctan\left[\frac{\bar{n}_k\sin(2\tilde{g}t\eta_k^2)}{\bar{n}_k+1-\bar{n}_k\cos(2\tilde{g}t\eta_k^2)} \right]}
\end{array}
\end{equation}
the approximated atomic inversion (\ref{appinv}) can be written as
\begin{equation}\label{atinv3}
 W(t)=\frac{\cos(2\tilde{g}t+\theta)}{r},
\end{equation}
where $\theta=\sum_k\theta_k$ and $r=\prod_kr_k$. In fig.
\ref{fig3}, five examples of the atomic inversion are displayed for
different powers $s$. Perfect revivals clearly occur for the $s=0$,
$s=1$ and $s=2$ cases, while for $s=-2$ and $s=-1$ the revivals are
never fully perfect even at long times. This is because if
$s=0,1,2,...$ we have
$\eta_k^2=\kappa\omega_k^s=\kappa\omega_0^sk^s\equiv\eta_0^2j$ where
$j$ is a positive integer, which does not hold if $s=...,-2,-1$.
Note that the revivals are a consequence of the non Markovian
treatment of the problem; the dynamics is unitary and each phonon
mode is considered at the same footing as the cavity mode and the QD
themselves. Therefore. also when a large number of different phonon
modes are coupled to the QD-cavity system exhibits revivals, however
less pronounced.

\section{Beyond the rotating wave approximation}\label{sec4}
So far all the results have been derived within the RWA and the
Lamb-Dicke approximation, which for $g\eta/\omega\ll1$  is expected
to be justified. In this section we calculate the atomic inversion
by numerically diagonalyze the truncated Hamiltonian of eq.
(\ref{noRWA}). The size of the Hamiltonian is chosen such that
convergence of its eigenstates is guaranteed. We restrict the
analysis to the one phonon mode case, which already explains most
effects. In the Fock state basis $\{|m\rangle\}$ of the phonon mode,
the matrix elements of the Hamiltonian are obtained by using the
formula
\begin{equation}\label{matrixel}
\begin{array}{lll}
 D_{mn} & \equiv & \langle m|\hat{D}(\eta)|n\rangle= \\ \\
& & \displaystyle{= e^{-|\eta|^2/2}\eta^{n-m}\!\sqrt{\!\frac{m!}{n!}}L_{m}^{n-m}\!(|\eta|^2), \hspace{0.5cm} m\leq n,}
\end{array}
\end{equation}
where $L_i^j$ is an associated Laguerre polynomial.

\begin{figure}[h]
\begin{center}
\includegraphics[width=8cm]{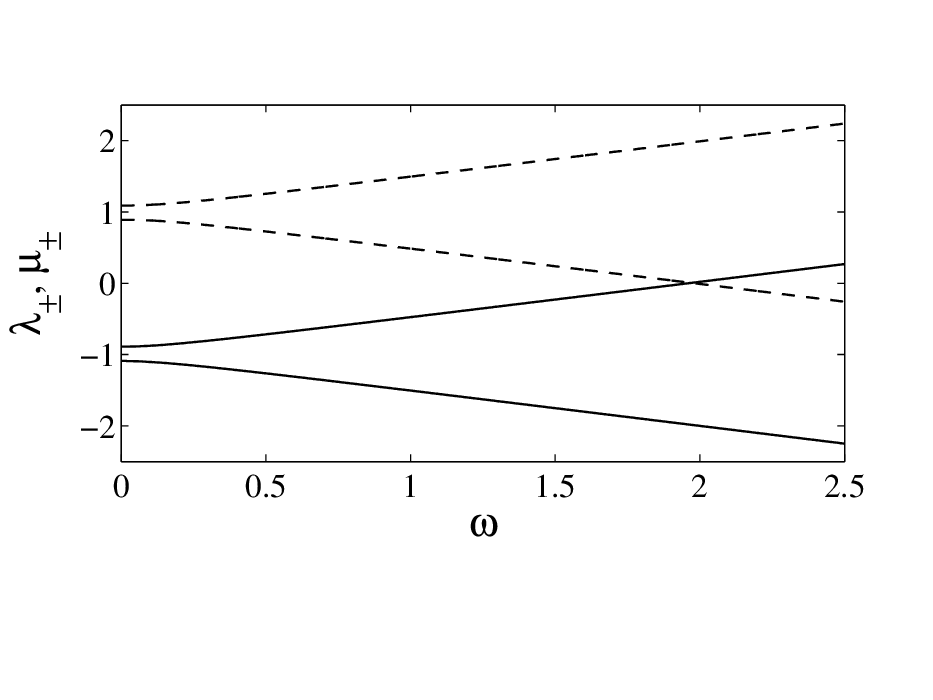}
\caption{\label{fig4} The eigenvalues (\ref{eigen}) as function of $\omega$, with $\eta^2=0.01$. Dashed curves show $\mu_\pm$ and solid curves $\lambda_\pm$. }
\end{center}
\end{figure}

\begin{figure}[h]
\begin{center}
\includegraphics[width=8cm]{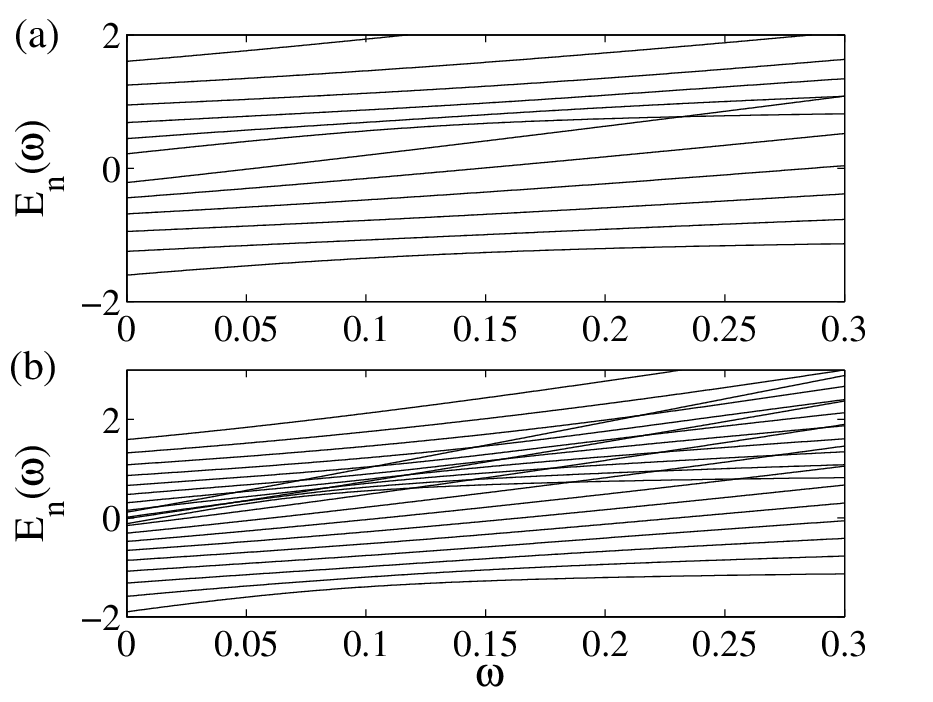}
\caption{\label{fig5} The numerically obtained eigenvalues as function of $\omega$, with $\eta^2=0.05$. In (a) the truncated Hamiltonian has size $12\times12$ and in (b) $22\times22$. Due to the non-RWA coupling the curve crossings become avoided, which is, however, not seen in the plots of this sice. }
\end{center}
\end{figure}

For lowest order truncation of the Hamiltonian, one keeps only a
single Fock state of  the phonon mode, and the RWA results given
above are achieved \cite{jonas}. In this approximation, when the
initial states of the phonon modes are on the form
$|c_{n_k}|=\delta_{n_k,n_k'}$, the combined quantum dot-cavity
system persists perfect Rabi oscillations with a rescaled vacuum
Rabi frequency $g\rightarrow
\tilde{g}\prod_{n_k'}L_{n_k'}(\eta_{n_k}^2)$. For zero temperature,
all the modes are in the vacuum and the above approximation simply
gives a rescaled frequency $g\rightarrow \tilde{g}$, which is the
result presented in \cite{chinese}. Thus, this simple derivation
regains the results of \cite{chinese}, and it is also clear how the
approxmation comes about and may be easily extended to non-zero
temperature phonon baths. A deeper insight of the approximation
(RWA) is gained by increasing the size of the truncated Hamiltonian.
In other words, to go beyond the RWA one needs to include more
coupling terms arising due to the phonon bath. Considering an
initial vacuum phonon mode and including first order corrections to
the RWA, the Hamiltonian can be written in matrix form (after an
overall shift in energy) as
\begin{equation}\label{appham}
 H_{vac}=\left[
\begin{array}{cccc}
\displaystyle{-\frac{\omega}{2}} & 0 & D_{00} & D_{01} \\ \\
0 & \displaystyle{\frac{\omega}{2}} & D_{01} & D_{11} \\ \\
D_{00} & D_{01} & -\displaystyle{\frac{\omega}{2}} & 0 \\ \\
D_{01} & D_{11} & 0 & \displaystyle{\frac{\omega}{2}}
\end{array}\right],
\end{equation}
with eigenvalues
\begin{equation}\label{eigen}
\begin{array}{l}
\lambda_\pm = \displaystyle{-\frac{D_{00}+D_{11}}{2}\!\pm\frac{1}{2}\!\sqrt{\left(D_{00}-D_{11}+\omega\right)^2+4D_{01}^2}}\\ \\
\mu_\pm = \displaystyle{\frac{D_{00}+D_{11}}{2}\!\pm\frac{1}{2}\!\sqrt{\left(D_{00}-D_{11}-\omega\right)^2+4D_{01}^2}}.
\end{array}
\end{equation}
These eigenvalues are shown in fig. \ref{fig4} as function of
$\omega$ for fixed $\eta$; $\lambda_\pm$ (solid lines) and $\mu_\pm$
(dashed lines). The {\it bare curves},
$\epsilon_\pm^{\lambda}=-D_{ii}\pm\omega/2$ and
$\epsilon_\pm^{\mu}=D_{ii}\pm\omega/2$, where $i=0,1$, are coupled
by the non-RWA term $D_{01}$. This makes the crossings at $\omega=0$
to be {\it avoided}, the degeneracies are lifted. Thus, in the RWA,
the difference in the values $D_{ii}$ for $i=0,1,2,...$ of the bare
energies at $\omega=0$ causes the collapse-revival structure.
However, as the non-RWA terms start to dominate the spectrum, the
bare energies are no longer proportional to $\pm\omega$ and the full
system dynamics will no longer show the nice collapse-revival
structure. For which $\omega$ that this non linear behaviour occurs,
depends on $g$ and $\eta$. A rough estimate  can be derived by
assuming that the linear dependence of $\omega/2$ should overrule
over the avoided crossing energy difference $2D_{01}$, which after
expanding in $\eta$ gives $\omega\gg4g\eta$. In other words, using
our typical parameter values of the previous section, we note that
when $\omega<1$ the RWA is likely to break down. For small $\omega$,
the eigen energies are very densely spaced, and a small non-RWA
correction will couple the different energies in an involved way.
The number of states taken into account to correctly describe the
dynamics is then growing rapidly, which can be seen in fig.
\ref{fig5} showing the numerically obtained eigenvalues for a
$12\times12$ dimensional (a) and a $22\times22$ dimensional (b)
Hamiltonian. Clearly the more states taken into account, the more
complicated energy spectrum.

\begin{figure}[h]
\begin{center}
\includegraphics[width=8cm]{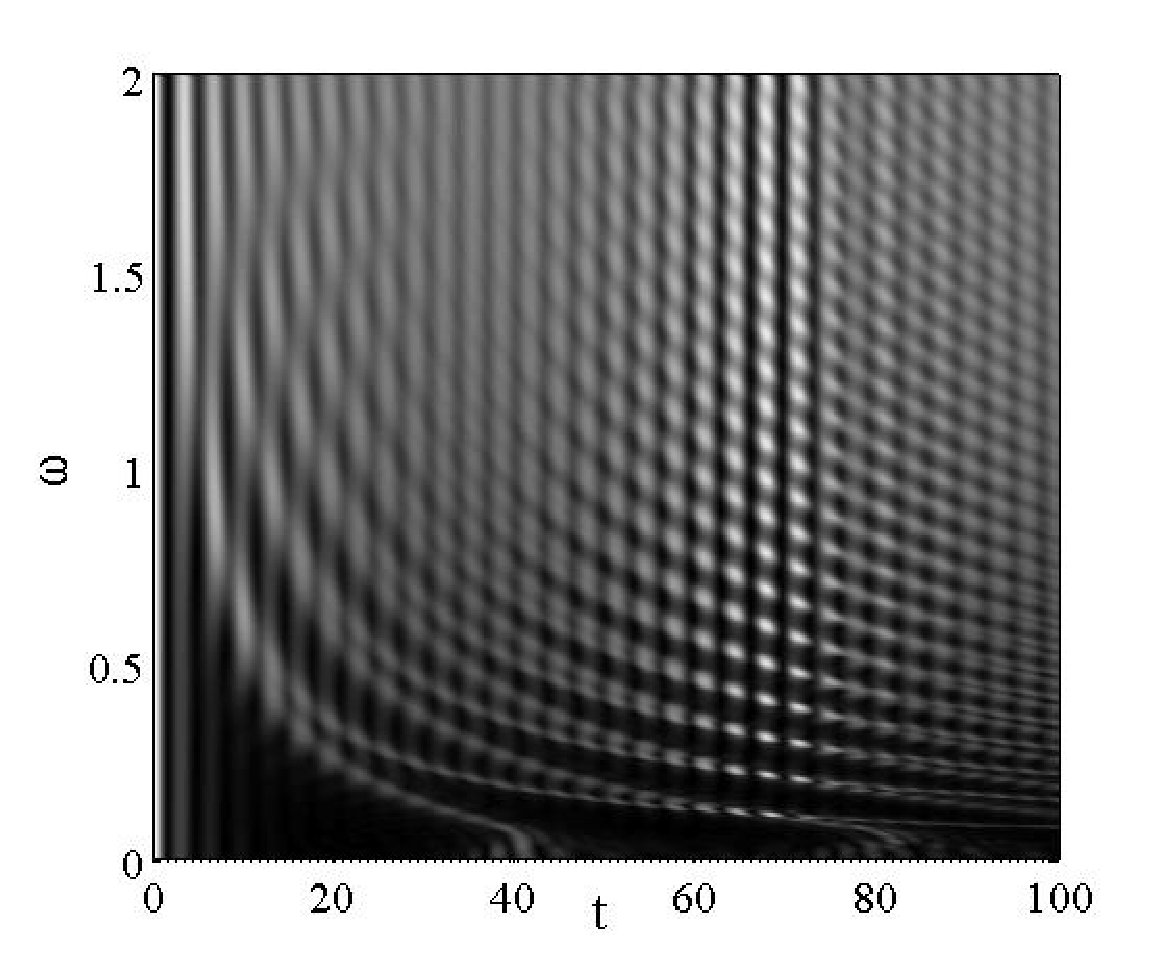}
\caption{\label{fig6} The inversion as function of $\omega$ in the regime where the RWA breaks down. The revivals at $t_{rev}=\pi/\eta^2$ build up for increasing $\omega$. Here $\eta^2=0.01$ and $\bar{n}=3$. }
\end{center}
\end{figure}

Finally, in fig. \ref{fig6} we give one example of the numerically
obtained inversion in a regime where the RWA is not justified.
Interestingly, for $\omega\ll1$ the state $|g\rangle$ (opposite of
the initial state) of the QD is more strongly populated, while for
increasing $\omega$ the collapse-revival pattern appears and
$W(t)\rightarrow1/2$ in the collapse regions.

\section{Conclusions}

We have studied a quantum dot strongly coupled to a single
high-finesse optical cavity mode by applying methods usually applied
in ion-laser interactions, namely the decomposition of the Glauber
displacement operator in Laguerre polynomials. This allowed us to
obtain results for the system when it is coupled to a phonon
reservoir beyond the Born-Markov approximation. We have studied
several cases, including $N$ identical and $N$ different phonon
modes, i.e. the case of non-zero temperature. Expressions for the
collapse and revival times of the Rabi oscillations have been
derived analytically, valid in a large range of parameters.

The analysis is carried out in the resonance condition
$\Delta=\Delta_\eta$, which in the RWA gives that no population is
transferred between the QD and the phonon reservoir. If this
resonance is not fulfilled, but other specific types of conditions
are (blue or red detuned), one may derive different types of
effective Hamiltonians in the RWA, in much resemblance with ion-trap
cavity systems \cite{plk}. Then the effective models will be
described by typical generalized two-mode Jaynes-Cummings
Hamiltonians, which can often be solved analytically \cite{jc}.
Another assumption made in the derivation is that the cavity mode is
initially in vacuum, while for a general initial state each photon
states will, just like in the case of the various phonon states,
induce different Rabi frequencies affecting the collapse-revival
pattern. Such situation may be of interest for state preparation or
state measurement and is left for future considerations.

We have also presented a short analysis of the dynamics in the non
RWA regime. In this parameter range, the energy spectrum becomes
very complex with crossing energy curves, and as the energy spacing
between the curves is small for these parameters, many eigenstates
of the Hamiltonian must be included in order to correctly describe
the dynamics. Non the less, from such an approach it is seen how the
RWA results are obtained as a first order correction to the trivial
situation, and how higher order terms cause avoided crossings
between the energies, and they are therefore no longer proportional
to $\pm\omega$ around the crossings which will induce a
``breakdown'' of the collapse-revival pattern. The higher order
terms in such an expansion can be seen as a virtual exchange of
phonons \cite{jonas2}.

\begin{acknowledgments}
This work was supported by EU-IP Programme SCALA (Contract No.
015714), the Swedish Government/Vetenskapsr\aa det, NORDITA and  Consejo
Nacional de Ciencia y Tecnolog\'{\i}a.
\end{acknowledgments}

\end{document}